%% file: main.tex
\PassOptionsToPackage{sort&compress}{natbib}
\documentclass[pdflatex,sn-nature]{sn-jnl}
\usepackage{setspace}
\usepackage{mathptmx}
\renewcommand{\normalsize}{\fontsize{12}{14.4}\selectfont}
\normalsize
\usepackage{graphicx}
\setcitestyle{super,open={},close={},comma}
\usepackage{multirow}
\usepackage{amsmath,amssymb,amsfonts}
\usepackage{amsthm}
\usepackage{xcolor}
\usepackage{textcomp}
\usepackage{bookmark}
\usepackage{manyfoot}
\usepackage{booktabs}
\usepackage{algorithm}
\usepackage{algorithmicx}
\usepackage{algpseudocode}
\usepackage{listings}
\theoremstyle{thmstyleone}

\theoremstyle{thmstyletwo}

\theoremstyle{thmstylethree}

\begin{document}
\title[AI Alignment Amplifies Hiring Discrimination]{AI Alignment Amplifies the Role of Race, Gender, and Disability in Hiring Decisions}
\author*[1, 2]{\fnm{Ze} \sur{Wang}}\email{ze.wang.19@ucl.ac.uk}

\author*[3]{\fnm{Guobin} \sur{Shen}}\email{shenguobin2021@ia.ac.cn}

\author[1]{\fnm{Michael} \sur{Thaler}}\email{michael.thaler@ucl.ac.uk}

\affil*[1]{\orgdiv{Department of Economics}, \orgname{University College London}, \orgaddress{\city{London}, \country{UK}}}
\affil[2]{\orgdiv{Stone Centre on Wealth Concentration, Inequality, and the Economy}, \orgname{University College London}, \orgaddress{\city{London}, \country{UK}}}
\affil[3]{\orgdiv{Institute of Automation}, \orgname{Chinese Academy of Sciences}, \orgaddress{\city{Beijing}, \country{China}}}

\maketitle

\noindent\textbf{Humans increasingly delegate consequential decisions to language models, yet whether these systems reproduce or reshape human patterns of discrimination remains unclear\cite{Obermeyer2019, Lambrecht2019, Caliskan2017, Hofmann2024, Lippens2024, wang-etal-2024-jobfair, An2025LLMbias, Gaebler2024, Glazko2024}. Here, across 29 models and 177 occupations covering nearly half of U.S. employment\cite{ONET2024, CPS2024, OES2024, BLSdisability2024}, we show that language models incorporate demographics into hiring decisions, advantaging female and Black candidates while penalising disabled candidates, with effect sizes comparable to six months to one year of additional education. While pre-trained models show small demographic effects, post-training alignment, which adapts models to human norms and preferences, amplifies advantages for female and Black candidates by 396\% and 413\% and worsens the disability penalty by 152\%. Compared with human employers in past correspondence experiments\cite{Bertrand2004, Quillian2017, Kline2022, Ameri2018, LIPPENS2023104315}, language models reverse racial discrimination, substantially attenuate the disability penalty, and amplify the female advantage. Investigating the mechanisms, we find behavioural patterns that parallel statistical discrimination in human labour markets, but disability consistently fares worst across all channels. We trace this asymmetry to the composition of alignment data, where disability is structurally underrepresented, and to models' internal representations, where alignment shifts the encoding of disability most negatively among the three marginalised groups.}

\section{Introduction}\label{sec1}
Artificial intelligence systems increasingly participate in decisions that were once exclusively human, from medical diagnoses~\cite{Esteva2017, McKinney2020} to judicial decisions~\cite{Kleinberg2018} to hiring~\cite{Lippens2024}, raising the concern that they may inherit the biases documented in decades of psychology and economics research~\cite{Caliskan2017, Hofmann2024}. Among the most consequential of these is discrimination based on demographic identity in labour markets~\cite{becker, 49fcf56e-096c-3297-8857-ef41b13d697d, arrow1971discrimination, Coffman2023, 10.1162/rest_a_01367, aigner1977statistical, 10.1257/aer.20171829, adc717f6-d16d-39e8-a957-19e5ff4ad9af}, where past correspondence experiments have consistently shown that racial and disability discrimination persist while gender discrimination has reversed into a slight female advantage~\cite{Bertrand2004, Quillian2017, Kline2022, Ameri2018, LIPPENS2023104315}. Whether language models reproduce these patterns, and whether alignment reshapes them, remains less known.

Previous studies have audited language models for hiring bias~\cite{Lippens2024, An2025LLMbias, wang-etal-2024-jobfair, Gaebler2024, Glazko2024}, establishing that discrimination can occur, but evidence is limited to surface patterns. One fundamental question is whether language models discriminate in ways that resemble human employers~\cite{Bertrand2004, Kline2022, LIPPENS2023104315}, who can penalise marginalised candidates based on group identity alone~\cite{becker, arrow1971discrimination} or instead favour them by inferring greater ability from the same credentials~\cite{aigner1977statistical, 10.1257/aer.20171829}, and whether such similarities in mechanisms lead to similarities also in hiring outcomes. Another is which part of the training process gives rise to the observed bias and why, a question that requires a framework to isolate the causal effect of alignment and to look inside the models to understand how they process demographic information.

\begin{figure}[t]
\centering
\includegraphics[width=1\textwidth]{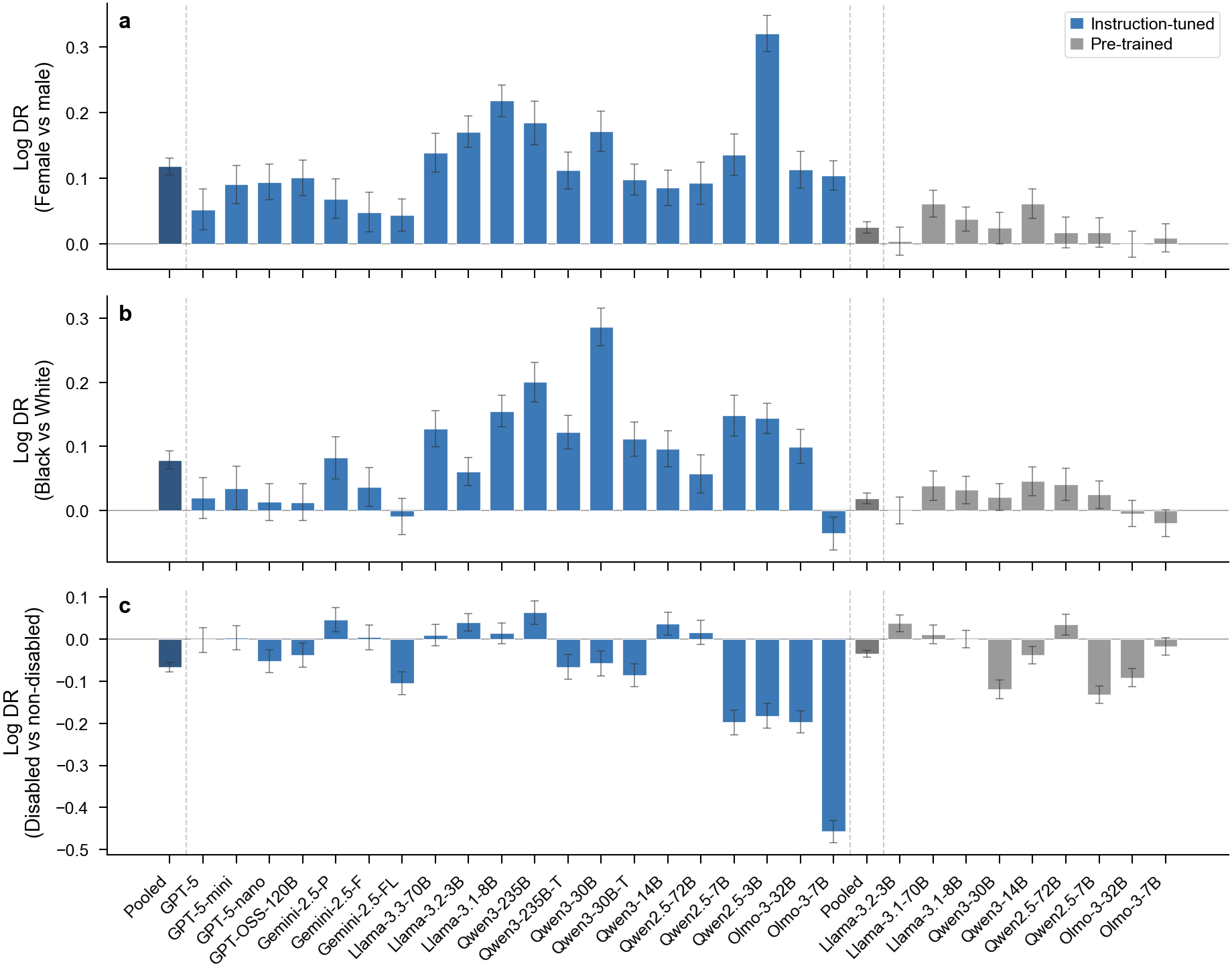}
\caption{\textbf{Demographic effects across language models.} Log discrimination ratios (i.e., the marginalised group's hiring rate relative to the non-marginalised group's) from per-model regressions for each of 29 models (Methods~III); positive values indicate an advantage for the marginalised group, negative values a penalty. Blue: instruction-tuned; grey: pre-trained; darker shading indicates pooled estimates (Extended Data Table~1). Error bars represent 95\% confidence intervals. \textbf{a}, Female vs.\ male. \textbf{b}, Black vs.\ White. \textbf{c}, Disabled vs.\ non-disabled.}\label{fig1_permodel}
\end{figure}

Here, we show, across 29 models and 177 occupations covering nearly half of U.S. employment~\cite{ONET2024, CPS2024, OES2024, BLSdisability2024} (Extended Data Fig.~1), that language models systematically incorporate demographics into hiring decisions, advantaging female and Black candidates while penalising disabled candidates (Fig.~1). Pre-trained models, which are trained on naturally occurring text, show small demographic effects, whereas instruction-tuned models, which are further trained on curated alignment data to conform to human norms and preferences, are far more sensitive to demographic identity. Compared with decades of correspondence experiments among humans~\cite{Bertrand2004, Quillian2017, Kline2022, Ameri2018, LIPPENS2023104315}, language models reverse racial discrimination, amplify the female advantage, and substantially attenuate, but do not eliminate, the disability penalty (Fig.~2). 

Yet the mechanisms behind these outcomes are familiar: models reward the same qualifications more generously for marginalised-group candidates~\cite{aigner1977statistical, 10.1257/aer.20171829, adc717f6-d16d-39e8-a957-19e5ff4ad9af} and penalise them more heavily when productivity signals are absent~\cite{49fcf56e-096c-3297-8857-ef41b13d697d, arrow1971discrimination, Coffman2023, 10.1162/rest_a_01367}. These patterns are weak in pre-trained models; alignment amplifies both channels, but not equally across groups: disabled candidates receive the smallest credential premium and the largest penalty.

We trace the asymmetry in alignment's effect to unequal representation of marginalised groups in the training data and to how alignment reshapes models' internal encoding of demographic information. Within the alignment data of Olmo-3, the only model family in our sample with fully open training data, disability is the least mentioned, has the least content addressing how disabled people are perceived or treated, and the gap widens further for content that advocates for the group's equality (Fig.~4a). The imbalance is reflected in how models process demographics internally: alignment shifts the internal representation of disability toward lower hiring odds while shifting gender and race in the opposite direction (Fig.~4b).

To our knowledge, this is the first study to investigate the mechanisms behind language models' hiring behaviour, and the first to trace these behavioural patterns to models' training process and internal representations. We introduce a framework that generates candidate profiles from public labour-market data, each with fully quantified qualifications and explicitly stated demographic attributes (Methods~I), enabling decomposition of each model's hiring decision into the weight it places on qualifications and demographics, thereby revealing the mechanisms behind its demographic treatment (Supplementary Information, `Derivations'). We show that the disparate treatment by language models departs from human discrimination by improving outcomes for historically marginalised groups while introducing new asymmetries between them, extending a literature and policy discourse focused on discrimination against protected groups. The effects we document are not artefacts of individual models or occupations: they appear across every model family, from 3 billion to 235 billion parameters, in open-weight and proprietary systems alike, and across 177 U.S. occupations, greatly expanding the empirical scope of prior work. Crucially, by comparing pre-trained and instruction-tuned models of the same architecture, we provide the first causal evidence that post-training alignment drives both demographic effects and asymmetries between historically marginalised groups.

\section{Asymmetry across marginalised groups}

The effects of demographic identity on hiring decisions are shown in Fig.~1. Every instruction-tuned model favours female candidates over otherwise identical male candidates, and all but two favour Black candidates. Disability shows a different pattern: the log discrimination ratios are heterogeneous in sign, and the average is significantly negative. Pre-trained models show the same directional asymmetry but with smaller magnitudes. Pooling across instruction-tuned models, the aggregate specification (Methods~III) yields significant advantages for female (log DR $+0.119$, $P < 0.001$; all reported $P$ values are two-sided) and Black ($+0.079$, $P < 0.001$) candidates but a significant penalty for disabled candidates ($-0.066$, $P < 0.001$; Extended Data Table~1), robust to nonparametric specification (Extended Data Tables~10--11). In relative terms, female candidates are selected 13\% more often, and Black candidates 8\% more often, than otherwise identical male and White candidates, effects comparable to roughly one year of additional education (13\%) for gender and eight months for race, while disabled candidates are selected 6\% less often.

These asymmetries emerge alongside a dramatic increase in qualification responsiveness: pre-trained models show minimal responsiveness to qualifications (pseudo~$R^2 \approx 0$), whereas instruction-tuned models are far more responsive (pseudo~$R^2 = 0.29$; Extended Data Table~1), with all qualification coefficients increasing three- to sixfold. To test whether alignment drives both, we compare ten matched instruct--base pairs sharing the same architecture. We find that alignment amplifies the female advantage by 396\% ($\Delta$ log DR $+0.101$, $P < 0.001$) and the Black advantage by 413\% ($+0.077$, $P < 0.001$), while worsening the disability penalty by 152\% ($-0.053$, $P < 0.001$; Extended Data Table~2). Simultaneously, it increases all qualification weights by 235--352\%, transforming pre-trained models into systems highly responsive to qualifications. These demographic effects are additive: pairwise interaction terms (female$\times$Black, female$\times$disabled, Black$\times$disabled) are not significant in either model type (Extended Data Table~3).

\section{Human comparison and occupation effects}

\begin{figure}[t]
\centering
\includegraphics[width=1\textwidth]{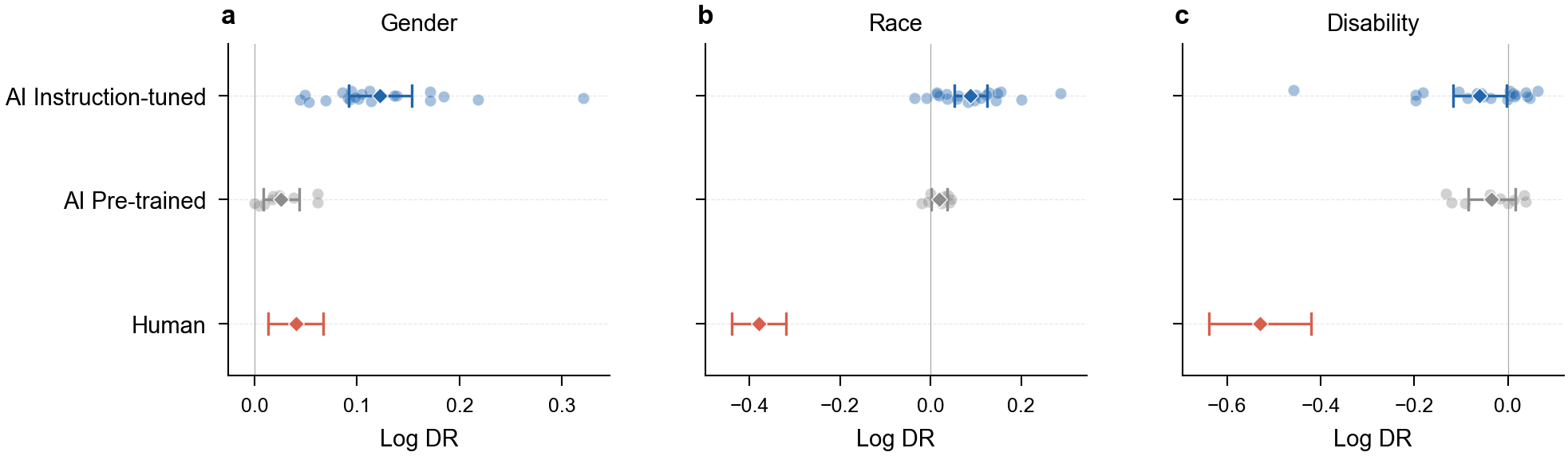}
\caption{\textbf{AI versus human hiring discrimination.} Meta-analytic estimates (diamonds) with 95\% confidence intervals for 20 instruction-tuned (blue) and 9 pre-trained (grey) models, and human correspondence study benchmarks (orange; Methods~IV). Each dot represents one model's log discrimination ratio; positive values indicate a marginalised-group advantage, negative values a penalty. \textbf{a}, Female vs.\ male. \textbf{b}, Black vs.\ White. \textbf{c}, Disabled vs.\ non-disabled.}\label{fig1}
\end{figure}

Despite learning from human-generated text, language models produce a markedly different pattern of discrimination~\cite{Bertrand2004, Quillian2017, Kline2022, Ameri2018, LIPPENS2023104315}. Aggregating per-model estimates by random-effects meta-analysis (Fig.~2; Methods~III), we find that for race, where equally qualified Black candidates receive 32\% fewer positive responses from human employers (log DR $-0.379$), language models reverse the direction and favour them by 9\% ($+0.088$). For disability, where humans give 41\% fewer positive responses ($-0.530$), AI attenuates the penalty to 6\% ($-0.060$) but does not reverse it. For gender, where humans show a slight 4\% female advantage ($+0.040$), language models amplify it to 13\% ($+0.123$).

\begin{figure}[t]
\centering
\includegraphics[width=\textwidth]{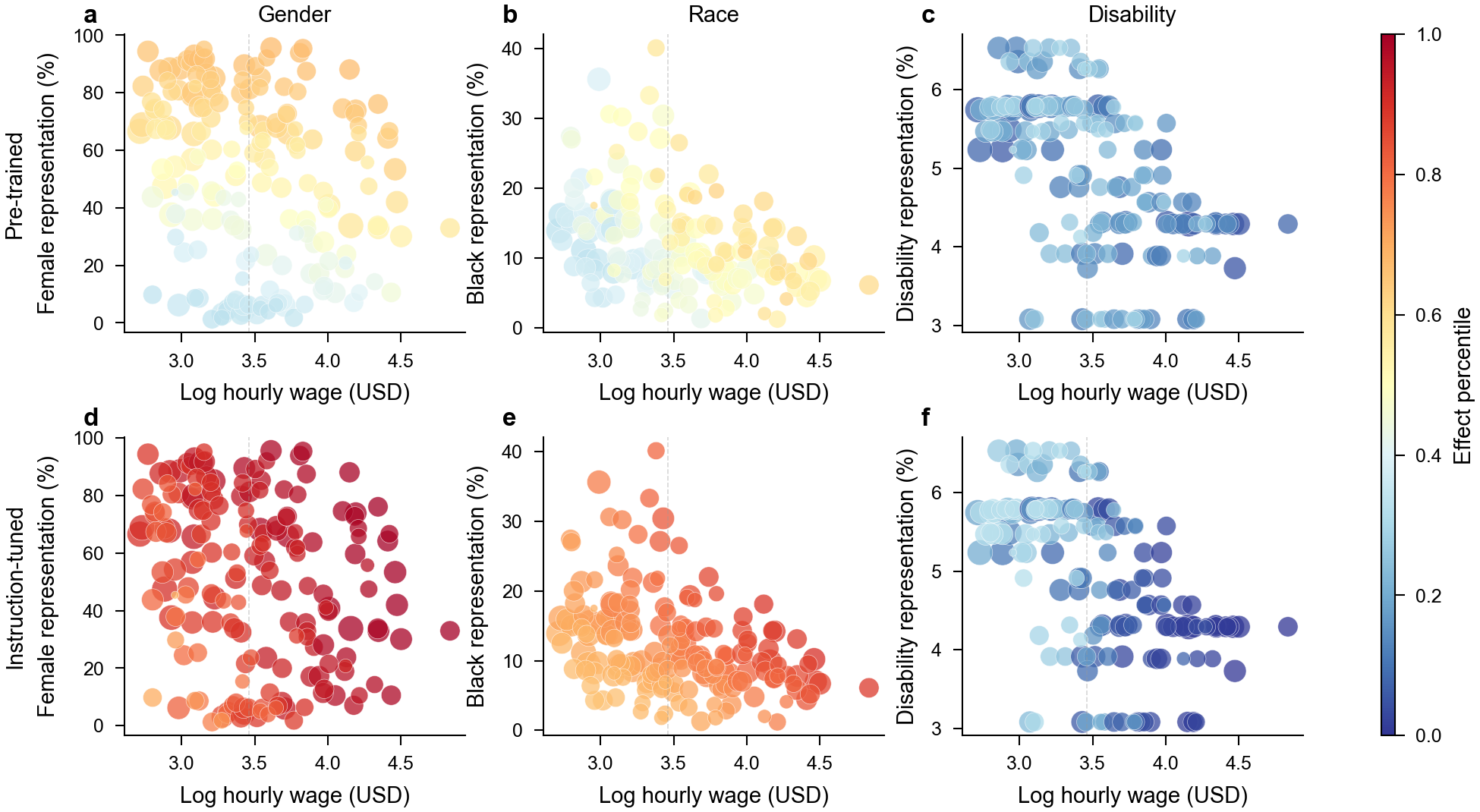}
\caption{\textbf{Occupation-level variation in demographic effects.} Each point represents one of 177 occupations, positioned by log hourly wage ($x$-axis) and group-specific representation rate ($y$-axis); point size proportional to log total employment. Colour indicates the fitted log discrimination ratio on a shared percentile scale across panels (red: higher; blue: lower). \textbf{a}--\textbf{c}, Pre-trained models. \textbf{d}--\textbf{f}, Instruction-tuned models. Columns: gender, race, and disability.}\label{fig6}
\end{figure}

These effects show some heterogeneity across the labour market. Interacting all regressors with occupation-level log hourly wage, total employment~\cite{OES2024}, and group-specific representation rates~\cite{CPS2024, BLSdisability2024} (Fig.~3), we find that the female advantage grows with log hourly wages (Female $\times$ Wage: $+0.023$, $P < 0.001$; Extended Data Table~4), and to a lesser extent with employment size (Female $\times$ Employment: $+0.011$, $P = 0.076$). The Black advantage shows a marginally significant wage gradient (Black $\times$ Wage: $+0.014$, $P = 0.051$). Disability, by contrast, is invariant to wage and employment size but shows a significant negative interaction with disability representation (Disabled $\times$ Rate: $-0.058$, $P = 0.020$).

The matched-pair design reveals that alignment causally introduces wage and employment-size gradients for gender ($\Delta$ Female $\times$ Wage: $+0.032$, $P < 0.001$; $\Delta$ Female $\times$ Employment: $+0.014$, $P = 0.039$) and a negative interaction between disability and disability representation ($\Delta$ Disabled $\times$ Rate: $-0.069$, $P = 0.016$), worsening the disability penalty where disabled workers are most prevalent. Beyond these gradients, alignment's demographic effects are largely uniform (Extended Data Table~5).

\section{Mechanisms of demographic effects}

The hiring outcomes differ, but the mechanisms do not. Models, like human employers~\cite{aigner1977statistical, 10.1257/aer.20171829, adc717f6-d16d-39e8-a957-19e5ff4ad9af}, reward the same qualifications more generously for marginalised-group candidates: all three marginalised groups receive higher returns to general skills, and female and Black candidates also receive a work-experience premium (Extended Data Table~6). Comparing matched instruct--base pairs, we find that alignment is a key driver behind this mechanism: alignment amplifies the return to general skills for all three groups, more than tripling it for female and Black candidates and more than doubling it for disabled candidates, and introduces a work-experience premium for female and Black candidates (Extended Data Table~7).

The parallel extends further. Human evaluators, when facing incomplete information, infer unknown attributes from demographic signals but hold lower expectations for marginalised groups~\cite{49fcf56e-096c-3297-8857-ef41b13d697d, arrow1971discrimination, Coffman2023, 10.1162/rest_a_01367}. Language models behave similarly: when work experience is absent, the general skills it would normally reveal are also hidden from the model (Methods~I), and models impose additional penalties on all three marginalised groups (Extended Data Table~8). Comparing matched instruct--base pairs, we find that alignment introduces these penalties: female ($\Delta$ log DR $-0.067$, $P < 0.001$), Black ($-0.054$, $P = 0.011$), and disabled ($-0.072$, $P < 0.001$; Extended Data Table~9).

To quantify the net contribution of the two mechanisms, we evaluate alignment-induced changes at the sample mean of qualifications (Supplementary Information, `Calculation'). Alignment rewards qualifications more for all marginalised groups, but the gain is smallest for disability: a candidate with average qualifications gains $+0.140$ log hiring odds if female, $+0.061$ if Black, but only $+0.037$ if disabled. At the same time, alignment increases the penalty when qualification signals are absent: $-0.014$ for female, $-0.011$ for Black, and $-0.015$ for disabled candidates. Thus, in both mechanisms, disability consistently receives the least favourable alignment effect, helping explain why alignment benefits female and Black candidates but not disabled candidates.
\section{Alignment data and representations}

\begin{figure}[t]
\centering
\includegraphics[width=1\textwidth]{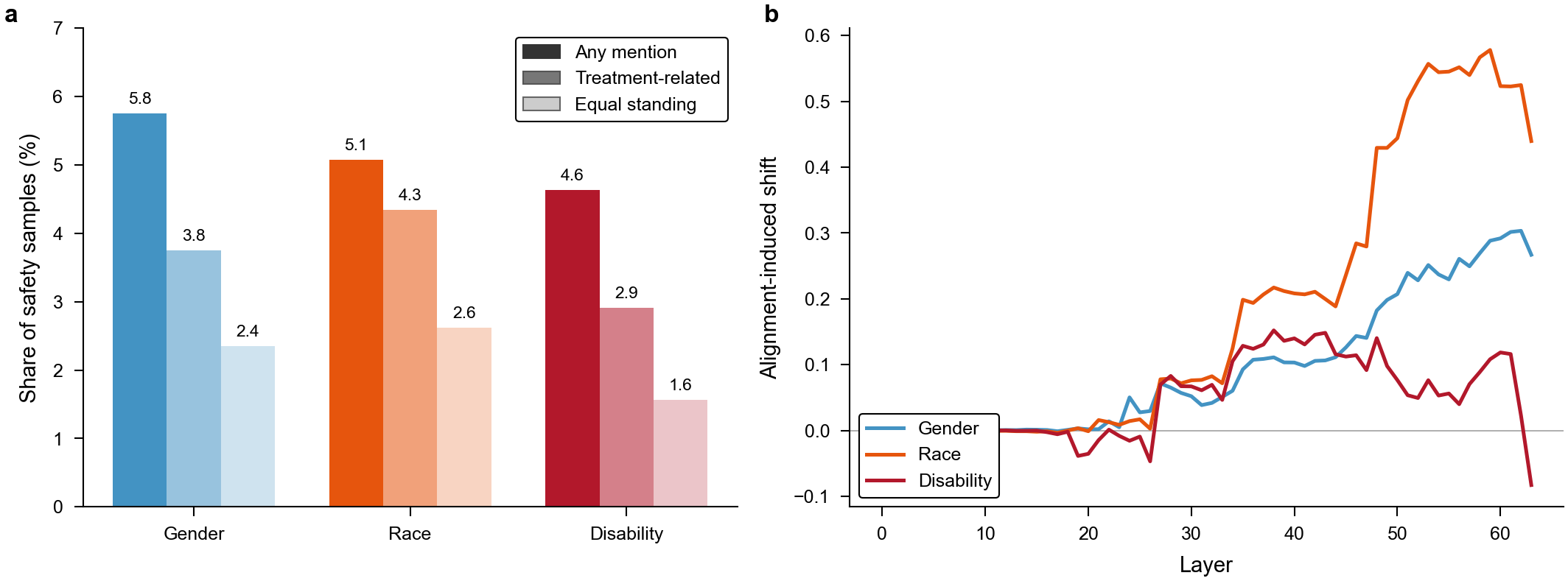}
\caption{\textbf{Alignment asymmetry in data and representations.} \textbf{a}, Share of Olmo~3 alignment training samples in the Safety domain of Dolci-Instruct-SFT ($n = 110{,}295$) that mention (any mention), engage with the perception or treatment of (treatment-related), or advocate for the equal standing of (equal standing) each demographic group, classified by an LLM judge (Methods~V). \textbf{b}, Alignment-induced shift in the contribution of each demographic to the hiring decision across transformer layers in Olmo-3-32B, measured as the difference in contrastive decision effect between the instruction-tuned and pre-trained models; negative values indicate that alignment suppresses hiring odds for the marginalised group (Supplementary Information).}\label{fig_repe}
\end{figure}

To investigate why alignment's effect differs for disability, we turn to the alignment data and models' internal representations. Within the safety domain of Olmo-3's alignment data, disability is consistently the least represented group, and the gap grows with the depth of engagement (Fig.~4a; Methods~V). At the surface level, disability is mentioned at roughly comparable rates to gender and race (4.6\% vs.\ 5.8\% and 5.1\%). But among content that addresses how the group is perceived or treated, disability falls 22\% below gender and 33\% below race. The gap widens further for content that advocates for the group's equality: here disability falls 34\% below gender and 40\% below race.

The imbalance in the training data is reflected in models' internal representations. Alignment reshapes how models encode demographic information at each transformer layer, measured as the difference in contrastive decision effect between the instruction-tuned and pre-trained models (Fig.~4b; Methods~V), where negative values indicate that alignment suppresses hiring odds for the marginalised group: in early layers, all three groups are near zero, but in deeper layers they diverge: gender and race become positive (reaching $+0.267$ and $+0.440$ at the final layer) while disability becomes negative ($-0.083$). Averaged across layers, disability shows the lowest shift ($+0.049$) relative to gender ($+0.094$) and race ($+0.184$), mirroring the behavioural asymmetry. We also test Olmo-3-7B and the largest matched pair, Qwen~2.5-72B; in both, the layer-averaged shift is again most negative for disability (Supplementary Information). 

\section{Discussion}

The key finding of this article is that alignment amplifies the role of demographic identity in language models' hiring decisions, but does not do so equally across historically marginalised groups. While pre-trained models show small demographic effects, alignment increases the advantages for female and Black candidates roughly fivefold while more than doubling the disadvantage for disabled candidates. This unequal amplification extends to occupation-level variation, with the female advantage growing in higher-wage occupations and the disability penalty deepening where disabled workers are most prevalent.

Why does alignment single out disability? There are many potential factors at play. One possibility is the historical gap between disability rights and other civil rights: the Civil Rights Act of 1964~\cite{CRA1964} prohibited discrimination based on race and sex but excluded disability, which did not receive comparable protection until the Americans with Disabilities Act of 1990 (ADA)~\cite{ADA1990}. Another is a research gap: correspondence experiments on disability remain far fewer than those on race or gender~\cite{LIPPENS2023104315}, and the AI fairness literature has also emphasised disability relatively less~\cite{Scully2025}.

A third factor is a conceptual mismatch. Anti-discrimination norms for race and gender centre on equal treatment: demographic identity should not influence decisions~\cite{CRA1964, NAS2004}. Disability poses a different challenge: the social model of disability~\cite{Oliver1990} holds that disadvantage arises not solely from impairment itself, but from the interaction between impairment and social and institutional barriers. Consistent with this perspective, the ADA framework emphasises reasonable accommodation rather than merely identical treatment~\cite{ADA1990}. However, empirically, evidence on the ADA's effects is mixed, with studies finding no improvement or even negative effects on the employment rates and wages of disabled workers, partly because accommodation requirements can raise the perceived cost of hiring~\cite{DeLeire2000, Acemoglu2001, BeegleStock2003, Hotchkiss2004}.

These factors find direct expression in the alignment data. Disability is the least represented group at every level of engagement, from surface mentions, to content addressing how the group is perceived, to content advocating for equality, and the gap widens at each successive level. The imbalance is reflected in the model's internal representations, where alignment shifts the encoding of disability in a direction that suppresses hiring odds far more than for gender or race.

The findings we document may extend beyond hiring. Where alignment data is domain-general, imbalances in the representation of demographic groups could lead to similar asymmetries in model behaviour across consequential decision contexts. Such compositional imbalances may also arise along other axes of identity, including personality, socioeconomic background, and language variety, and could plausibly produce analogous effects.

On the whole, language models exhibit distinct patterns of demographic effects compared with human employers, for both pre-trained and instruction-tuned models. Where human racial and disability discrimination have shown no noticeable improvement over decades of correspondence studies~\cite{Bertrand2004, Quillian2017, Kline2022, Ameri2018, LIPPENS2023104315}, language models reverse racial discrimination and attenuate the disability penalty by an order of magnitude, yet amplify the female advantage well beyond what has been documented for human employers. Current regulations such as the EU AI Act~\cite{EUAIAct2024} and New York City's Local Law 144~\cite{NYCLocalLaw144} focus on detecting aggregate discrimination against protected groups. Our findings point to concerns that go beyond this scope. First, AI-driven hiring creates new asymmetries within marginalised groups. Second, language models exhibit behaviour consistent with statistical discrimination in human labour markets~\cite{aigner1977statistical, 10.1257/aer.20171829, adc717f6-d16d-39e8-a957-19e5ff4ad9af, 49fcf56e-096c-3297-8857-ef41b13d697d, arrow1971discrimination, Coffman2023}, but whether their inferences from demographics reflect accurate or inaccurate beliefs about group differences remains open~\cite{10.1162/rest_a_01367}. This distinction matters: what we observe could be rational inference or harmful bias. Third, a structural concern: human hiring discrimination, though persistent, is concentrated among a minority of employers~\cite{Kline2022} and can in principle be addressed firm by firm. The patterns we document, by contrast, are robust across every model vendor and architecture tested. There is therefore a realistic possibility that as language models become standard tools in hiring, the asymmetries within marginalised groups will be replicated at scale across the labour market.

\bibliography{sn-bibliography}

\clearpage
\section*{Methods}\label{sec:methods}
\subsection*{I. Audit design}

\textit{Overview.} We construct a pipeline that generates occupation-specific candidate pools and role requirements from publicly available data~\cite{ONET2024, CPS2024, OES2024, BLSdisability2024}, with each candidate fully quantified as a vector of real-valued attributes. Candidates are paired so that they differ in both qualifications and demographics, and the quantified attributes enter directly as regressors. This enables multi-candidate comparisons at scale across the labour market.

\textit{Occupation sample.} We draw occupation-level data from four public U.S. sources: the O*NET 29.2 database~\cite{ONET2024} (required education, work experience, general skills, and technology skills), the 2024 Current Population Survey~\cite{CPS2024} (the gender and racial composition of each occupation), the May 2024 Occupational Employment and Wage Statistics~\cite{OES2024} (hourly wages and total employment), and the 2025 Labor Force Characteristics report~\cite{BLSdisability2024} (disability prevalence by occupation). Disability statistics are reported at the mid-level occupational category and mapped to detailed occupations via CPS classifications. We retain the intersection of occupations present in all four sources with non-suppressed values, yielding a final sample of 177 occupations with complete information on skills, education, experience, wages, and demographic composition.

\textit{Candidate profiles.} For each occupation, the O*NET database provides the five most important general skills (ranked by survey-based importance ratings), distributions of required education and work experience, and in-demand technology skills. We generate candidate profiles by sampling independently from these occupation-specific distributions. Each general skill level is drawn from a normal distribution centred on the occupation’s entry-level requirement and bounded at 0--7, with a common coefficient of variation ($= 0.24$) calibrated to the dispersion observed in U.S. standardised test scores~\cite{SAT2024, ACT2025} to prevent unrealistic concentration or fat tails in the general skill distribution. Education and work experience are sampled from the O*NET distributions, with the ``Less than a High School Diploma’’ category excluded and its probability redistributed, as this level may signal that a candidate is underage, triggering a refusal to evaluate rather than revealing a hiring preference. GPA is drawn from a normal distribution with the same coefficient of variation, mean $= 3.0$ for high school graduates~\cite{NAEP2009} and $3.15$ for college graduates~\cite{gradeinflation}, bounded at 0--4. For occupations with in-demand technology skills, candidates are assigned the full set with 50\% probability and none otherwise; for occupations without such requirements, no technology skills are assigned.

\textit{Resume construction.} Each candidate profile is converted into a natural-language resume structured to resemble a realistic CV, with an education section listing degree field and dates, and a work experience section containing dated employment periods and skill-based bullet points for each position. The conversion proceeds in two stages. In the first stage, GPT-4o generates two occupation-level components, each verified manually: (a) a mapping from each education level to the most relevant degree field, and (b) a career ladder of four job titles spanning entry-level to senior positions along the same occupational track. In the second stage, the candidate’s attributes are assembled into a formatted resume: the education section lists degree field and corresponding dates; the number of work experience positions is determined by years of experience (1 if $< 2.5$ years, 2 if $< 6$, 3 if $< 10.5$, 4 otherwise), job titles are drawn from the career ladder in descending seniority, and employment periods are computed backward from a fixed application date. For candidates with work experience, GPT-4o generates 3--4 bullet-point descriptions per position, conditioned on the candidate’s general skill names, levels, and occupation, with prompts enforcing a neutral style with no names, pronouns, or demographic references (Supplementary Information). For candidates without work experience, the work-experience section is omitted entirely; because general skills are embedded in the bullet points, they become unobservable to the model, providing the variation exploited in our analysis of missing qualification signals.

\textit{Demographic counterfactuals.} Each resume is replicated into eight variants spanning all $2 \times 2 \times 2$ combinations of gender (male, female), race (White, Black or African American), and disability status (reports a disability, reports no disability). A standardised demographic information block is inserted immediately after the Professional Objective section of each variant, stating the candidate's gender, race/ethnicity, and disability status alongside a Right to Work declaration (Supplementary Information).

\textit{Pairwise matching.} For each occupation, we generate 100 candidate profiles, yielding $100 \times 8 = 800$ demographic variants. We then randomly sample 400 pairs per occupation from the 800 demographic variants, requiring that each pair draws from two distinct candidate profiles, ensuring that paired candidates differ in qualifications. This yields $177 \times 400 = 70{,}800$ pairwise comparisons per model.

\textit{Hiring task.} Each model receives a three-part prompt for every pairwise comparison: (1) occupation-specific role requirements, extracted from O*NET data and rewritten by GPT-4o to produce professional language; (2) two candidate resumes, presented in randomised order; and (3) selection instructions asking the model to output ``1’’ or ``2’’ to indicate its choice. All prompt templates and resume examples are provided in the Supplementary Information.

\subsection*{II. Models}
We evaluate 29 language models. The 20 instruction-tuned models are: OpenAI GPT~\cite{GPT5} (GPT-5, GPT-5-mini, GPT-5-nano, GPT-OSS-120B), Google Gemini~\cite{Gemini25} (Gemini 2.5 Pro, 2.5 Flash, 2.5 Flash-Lite), Meta Llama~\cite{Llama3} (Llama 3.3-70B, 3.2-3B, 3.1-8B), Alibaba Qwen~\cite{Qwen25, Qwen3} (Qwen3-235B Instruct, Qwen3-235B Thinking, Qwen3-30B Instruct, Qwen3-30B Thinking, Qwen3-14B, Qwen2.5-72B, Qwen2.5-7B, Qwen2.5-3B), and AI2 Olmo~\cite{OLMo3} (Olmo-3-32B, Olmo-3-7B). The 9 pre-trained models are: Meta Llama (Llama 3.2-3B, 3.1-70B, 3.1-8B), Alibaba Qwen (Qwen3-30B, Qwen3-14B, Qwen2.5-72B, Qwen2.5-7B), and AI2 Olmo (Olmo-3-32B, Olmo-3-7B). We pair each Llama, Qwen, and Olmo instruction-tuned model with its pre-trained version where a publicly available checkpoint exists, excluding Qwen2.5-3B whose pre-trained model required excessive retries to yield valid responses. The resulting ten matched pairs enable within-architecture identification of the causal effect of alignment. The pre-trained version of Llama 3.3-70B is Llama 3.1-70B, and the Qwen3-30B Instruct and Thinking models share a single pre-trained version, Qwen3-30B; for the remaining pairs the pre-trained model carries the same name as its instruction-tuned counterpart. The exact model identifiers are listed in Supplementary Information (`Models'). All models are run with default inference settings.

\subsection*{III. Statistical analysis}

\textit{Aggregate specification.} Each pairwise hiring decision produces a binary outcome. We regress the log odds of selecting candidate $i$ over $j$ on the differences in their characteristics:
\begin{equation}\label{eq:original}
\log \frac{P(\text{select } i)}{P(\text{select } j)} = \sum_{d} \beta_d \, \Delta d_{ij} + X_{ij}^{\prime}\delta + \lambda \, \Delta p_{ij},
\end{equation}
where $\beta_d$ captures the hiring advantage (positive) or penalty (negative) for a marginalised-group candidate on demographic dimension $d$ relative to a majority-group opponent, holding qualifications $X$ and presentation order constant (Supplementary Information, `Derivations'). Because each decision is a binary choice, the odds of selecting one candidate over the other equal the ratio of the rates at which the two are selected, so $\beta_d$ measures the log discrimination ratio. GPA and the average of the five general skills (skill avg.) are $z$-scored before differencing; education (in years), work experience (in years), and technology skills (binary; zero for occupations without technical requirements) enter as raw differences. Because the model is specified in within-pair differences, the intercept cancels and is omitted. We report McFadden's pseudo-$R^2$ as the logit goodness-of-fit measure. All regressions cluster standard errors at the occupation level (177 clusters), with degrees of freedom set to $G_{\text{occ}} - 1 = 176$. We do not cluster at the model level because the number of models (9 pre-trained, 20 instruction-tuned) is too small for reliable cluster-robust inference~\cite{Cameron2015}; between-model variation is instead addressed through the random-effects meta-analysis reported in Fig.~2. As a robustness check, we also estimate two-way cluster-robust standard errors (occupation $\times$ model) following Cameron, Gelbach \& Miller~\cite{Cameron2011}; our main results for the asymmetry between gender, race, and disability are robust to this specification. Dropping Olmo-3-7B (both instruction-tuned and pre-trained), which shows the largest disability effect in our sample, likewise preserves the asymmetry. Coefficients are in log-odds units; to convert to the percentage change in odds, compute $(e^{\beta} - 1) \times 100\%$.

\textit{Cross-occupation variation.} To test whether demographic effects vary across the labour market, we interact each demographic indicator with three occupation-level characteristics: log hourly wage ($z$-scored; from OES), total employment ($z$-scored; from OES), and group-specific representation rate (percentile rank; from CPS and BLS). A positive interaction indicates that the demographic advantage or penalty is amplified in occupations where the characteristic is higher. This specification pools all observations across instruction-tuned or pre-trained models separately.

\textit{Mechanisms investigation.} We extend the aggregate specification (equation~\ref{eq:original}) to investigate two potential mechanisms. First, we add interactions between each qualification and each demographic group to test whether the returns to qualifications differ by group (differential returns). Second, we add an indicator $v = \mathbf{1}[\text{work exp.} = 0]$ and its interactions with each demographic group, testing whether models penalise candidates differently by group when qualification signals are absent (information asymmetry). We further include $v$'s interactions with qualifications and a conditional skill average $(1 - v) \times \overline{\text{skill}}$ to control for the reweighting of remaining signals when work experience is absent. Each specification pools all observations across instruction-tuned or pre-trained models separately.

\textit{Alignment effects.} To identify the causal effect of alignment, we restrict to the ten matched instruct--base pairs and pool their observations, interacting every regressor with an instruction-tuning indicator. The coefficient on each $\times$instruct interaction estimates the causal effect of alignment on that variable, holding model architecture fixed. We apply this approach to three specifications: the aggregate specification (equation~\ref{eq:original}), identifying the effect of alignment on demographic preferences; the cross-occupation specification, testing whether alignment introduces occupation-level gradients in demographic effects; and the mechanism specifications, testing whether alignment amplifies differential returns and information-asymmetry penalties.

\textit{Meta-analysis.} Equation~\ref{eq:original} is estimated separately for each model; the resulting demographic coefficients are then aggregated separately for the instruction-tuned and pre-trained groups using DerSimonian--Laird random-effects meta-analysis~\cite{DerSimonian1986}, which treats each model's estimate of a given demographic coefficient as a draw from a distribution of true effects and estimates both the mean and between-model heterogeneity. Because the number of models is small (20 instruction-tuned, 9 pre-trained), we apply the Hartung--Knapp adjustment~\cite{Hartung2001}, replacing standard normal critical values with $t(K-1)$ quantiles for more conservative inference, where $K$ is the number of models in each group.

\textit{Nonparametric robustness.} To test whether nonlinearity or higher-order interaction effects affect our conclusions, we estimate XGBoost~\cite{Chen2016XGBoost} (gradient-boosted trees; 300 rounds, max depth 6, learning rate 0.1, subsample 0.8) and random forest~\cite{Breiman2001} (500 trees, minimum leaf size 20) classifiers using raw candidate-level features $(X_i, X_j)$ rather than pairwise differences. Log discrimination ratios are computed via counterfactual prediction: for each pair, we flip a demographic indicator while holding all other features fixed and average the log-odds difference across pairs. Five-fold cross-validation confirms no overfitting (relative train--test logloss gap $< 5$\% for XGBoost, $< 15$\% for random forest). For inference, we use an occupation-level cluster bootstrap ($B = 1{,}000$)~\cite{Cameron2008}, resampling occupations with replacement and retraining on each bootstrap sample.

\subsection*{IV. Human study benchmarks}

The primary human benchmark is Lippens et al.~\cite{LIPPENS2023104315}, a meta-analysis of 169 correspondence experiments (2005--2020) across multiple countries. Reported discrimination ratios (DR): female $= 1.04$ (95\% CI 1.01--1.07; log DR $= 0.04$), Black $= 0.68$ (0.64--0.73; log DR $= -0.38$), disability $= 0.59$ (0.53--0.66; log DR $= -0.53$).

To verify robustness to benchmark choice, we also consider three further studies. For race, Quillian et al.~\cite{Quillian2017} meta-analyse 24 U.S. field experiments (1989--2015) and report that White applicants receive 36\% more callbacks than equally qualified African Americans (log DR~$\approx -0.31$). Kline et al.~\cite{Kline2022} document racial discrimination in a large-scale correspondence experiment across Fortune~500 firms (log DR~$\approx -0.09$). For disability, Ameri et al.~\cite{Ameri2018} find that applicants disclosing a disability receive 26\% fewer expressions of employer interest (log DR~$\approx -0.30$).

\subsection*{V. Alignment data and representations}

\textit{Alignment data composition.} Olmo-3~\cite{OLMo3} is post-trained on Dolci-Instruct-SFT ($n = 2{,}152{,}112$ samples). Of these, $110{,}295$ belong to the Safety domain (sourced from WildJailbreak~\cite{WildJailbreak2024}, WildGuardMix~\cite{WildGuardMix2024}, and CoCoNot~\cite{CoCoNot2024}). We classify each Safety sample by relevance to gender, race, or disability using an LLM judge under three progressively stricter definitions. The any-mention tier (GPT-5.4-nano) counts any reference to a demographic group; the treatment-related tier (GPT-5.4) requires the message to relate to how people from the group are or should be perceived, treated, or evaluated; the equal-standing tier (GPT-5.4) asks whether the conversation advocates for the equal standing of the group. Full prompt templates are provided in the Supplementary Information.

We validated all three classifiers against human annotations: the any-mention classifier achieves 98.2\% overall agreement (300 randomly sampled Safety conversations, annotated on all three dimensions; gender 98.3\%, race 98.0\%, disability 98.3\%); the treatment-related classifier achieves 99.0\% (300 labels from any-mention positives, 100 per dimension; gender 98\%, race 100\%, disability 99\%); and the equal-standing classifier achieves 98.7\% (300 labels from treatment-related positives, 100 per dimension; gender 97\%, race 100\%, disability 99\%).

\textit{Representation engineering.} For each candidate pair, both candidates begin at the baseline demographic profile (male, White, non-disabled); we then flip the first candidate's label on a single demographic dimension (e.g., to disabled), producing a baseline prompt and a counterfactual, with all qualification attributes fixed. At each transformer layer, we extract the last-token hidden states from both prompts, project the hidden-state difference onto the model's decision axis (i.e., the difference between the unembedding vectors for output tokens ``1'' and ``2'') to obtain the contrastive decision effect (CE) for that pair, and average over 1,770 randomly sampled pairs (10 per occupation); positive values indicate that the demographic label increases the model's tendency to select the first candidate. The alignment-induced shift is the difference in mean CE between instruction-tuned and pre-trained models. Full details are provided in the Supplementary Information.

\section*{Data Availability}
All datasets used in this study are publicly available. The O*NET 29.2 database~\cite{ONET2024}, produced by the U.S. Department of Labor, Employment and Training Administration and used under the CC~BY~4.0 license, can be found at \url{https://www.onetcenter.org/db_releases.html}. The Current Population Survey~\cite{CPS2024} can be found at \url{https://www.bls.gov/cps/cps_aa2024.htm}. The Occupational Employment and Wage Statistics~\cite{OES2024} can be found at \url{https://www.bls.gov/oes/tables.htm}. The Labor Force Characteristics report~\cite{BLSdisability2024} can be found at \url{https://www.bls.gov/news.release/disabl.htm}. The Dolci-Instruct-SFT alignment dataset~\cite{OLMo3} can be found at \url{https://huggingface.co/datasets/allenai/Dolci-Instruct-SFT}. The processed occupation dataset, generated candidate resumes and role requirements, all model responses, representation engineering outputs, and statistical analysis outputs are available in figshare (public upon publication).

\section*{Code Availability}
All code is available at Code Ocean (public upon publication).

\section*{Acknowledgements}
We thank M. Vera-Hern\'andez for feedback on an earlier version of the article.

\section*{Funding}
This work was funded by the Stone Scholar Grant at the Stone Centre at UCL.

\section*{Author Contributions}
Z.W., G.S., and M.T.\ designed the research. Z.W.\ and G.S.\ performed the experiments. Z.W.\ analysed the data and wrote the paper.

\section*{Competing Interests}
The authors declare no competing interests.

\section*{Additional Information}
Supplementary Information is available for this paper. Correspondence and requests for materials should be addressed to Z.W.\ or G.S.

\clearpage
\section*{Extended Data}\label{sec:extdata}
\setcounter{figure}{0}
\renewcommand{\figurename}{Extended Data Fig.}
\setcounter{table}{0}
\renewcommand{\tablename}{Extended Data Table}

\begin{figure}[h]
\centering
\includegraphics[width=1\textwidth]{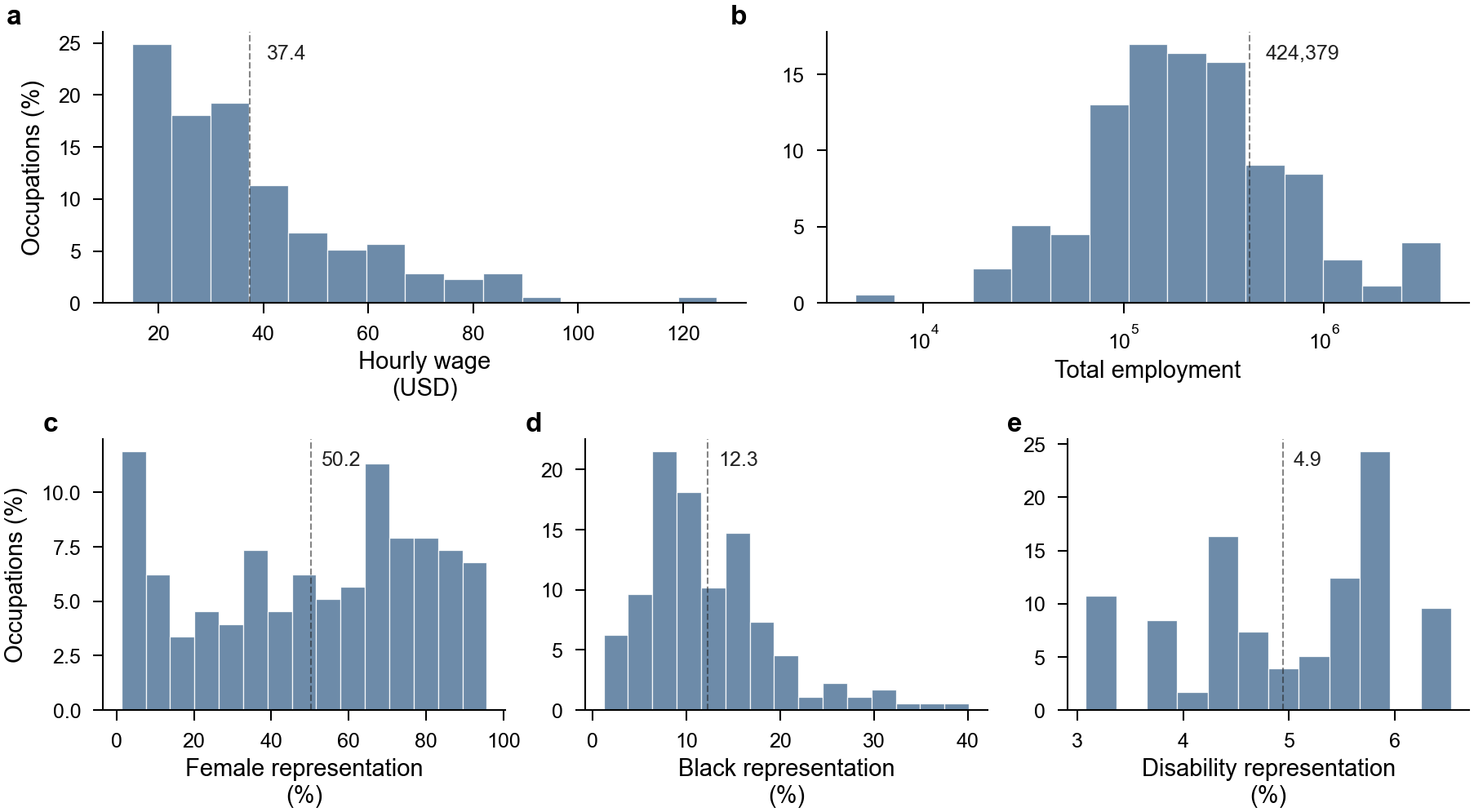}
\caption{\textbf{Distribution of occupation characteristics.} Histograms showing the distribution of (\textbf{a}) hourly wage, (\textbf{b}) total employment, (\textbf{c}) female representation rate, (\textbf{d}) Black representation rate, and (\textbf{e}) disability representation rate across 177 occupations. Dashed vertical lines indicate sample means. Employment is displayed on a logarithmic scale.}\label{figOccupation}
\end{figure}
\clearpage

\setcounter{table}{0}
\renewcommand{\thetable}{\arabic{table}}
\input{table_aggregate}
\clearpage
\input{table_aggregate_alignment}
\clearpage
\input{table_intersectional_aggregate}
\clearpage
\input{table_cross_occupation}
\clearpage
\input{table_cross_occupation_alignment}
\clearpage
\input{table_differential_returns}
\clearpage
\input{table_differential_returns_alignment}
\clearpage
\input{table_information_asymmetry}
\clearpage
\input{table_information_asymmetry_alignment}
\clearpage
\input{table_nonparametric_base}
\clearpage
\input{table_nonparametric_instruct}

\end{document}

%% file: table_aggregate.tex
\begin{table}[h!]
\centering
\parbox{0.85\textwidth}{\caption{\textbf{\textbar{}\ Aggregate Specification}}\label{tab:aggregate}}
\scriptsize
\begin{tabular}{lcc}
\toprule
 & (1) Base & (2) Instruct \\
\midrule
\textit{Panel A: Demographics} & & \\
Female & $0.0255^{***}$ $(0.0044)$ & $0.1186^{***}$ $(0.0065)$ \\
Black & $0.0187^{***}$ $(0.0044)$ & $0.0788^{***}$ $(0.0071)$ \\
Disabled & $-0.0345^{***}$ $(0.0042)$ & $-0.0663^{***}$ $(0.0059)$ \\[2pt]
\textit{Panel B: Controls} & & \\
Education & $0.0214^{***}$ $(0.0028)$ & $0.1220^{***}$ $(0.0116)$ \\
Work experience & $0.0939^{***}$ $(0.0028)$ & $0.5223^{***}$ $(0.0162)$ \\
GPA & $0.1422^{***}$ $(0.0030)$ & $0.4665^{***}$ $(0.0108)$ \\
Technology skills & $0.0881^{***}$ $(0.0071)$ & $0.4794^{***}$ $(0.0245)$ \\
Skill avg. & $0.0741^{***}$ $(0.0062)$ & $0.3386^{***}$ $(0.0182)$ \\
Position & $-0.0023$ $(0.0031)$ & $-0.0068$ $(0.0044)$ \\
\midrule
Observations & $637{,}200$ & $1{,}416{,}000$ \\
Pseudo $R^2$ & $-0.0030$ & $0.2886$ \\
\bottomrule
\end{tabular}
\vspace{2mm}
\parbox{0.745\textwidth}{\scriptsize\textit{Notes:} $^{***}p<0.01$, $^{**}p<0.05$, $^{*}p<0.10$. Logit on pairwise candidate differences; occupation-clustered standard errors in parentheses (177 clusters). All variables are differences between candidates $i$ and $j$ (e.g.\ Female $= \text{female}_i - \text{female}_j$). Each model evaluates 70{,}800 pairwise hiring decisions (177 occupations $\times$ 400 pairs). Column~(1): 9 pre-trained models ($9 \times 70{,}800 = 637{,}200$). Column~(2): 20 instruction-tuned models ($20 \times 70{,}800 = 1{,}416{,}000$).}
\end{table}

%% file: table_aggregate_alignment.tex
\begin{table}[h!]
\centering
\parbox{0.85\textwidth}{\caption{\textbf{\textbar{}\ Alignment Effects: Aggregate Specification}}\label{tab:aggregate_align}}
\scriptsize
\begin{tabular}{lcc}
\toprule
 & (1) Base & (2) $\times$ Instruct \\
\midrule
\textit{Panel A: Demographics} & & \\
Female & $0.0255^{***}$ $(0.0044)$ & $0.1008^{***}$ $(0.0072)$ \\
Black & $0.0187^{***}$ $(0.0044)$ & $0.0773^{***}$ $(0.0080)$ \\
Disabled & $-0.0345^{***}$ $(0.0042)$ & $-0.0526^{***}$ $(0.0070)$ \\[2pt]
\textit{Panel B: Controls} & & \\
Education & $0.0214^{***}$ $(0.0028)$ & $0.0753^{***}$ $(0.0085)$ \\
Work experience & $0.0938^{***}$ $(0.0028)$ & $0.3056^{***}$ $(0.0103)$ \\
GPA & $0.1422^{***}$ $(0.0030)$ & $0.3542^{***}$ $(0.0084)$ \\
Technology skills & $0.0881^{***}$ $(0.0071)$ & $0.2072^{***}$ $(0.0165)$ \\
Skill avg. & $0.0741^{***}$ $(0.0062)$ & $0.2300^{***}$ $(0.0141)$ \\
Position & $-0.0022$ $(0.0031)$ & $-0.0058$ $(0.0047)$ \\
\midrule
Observations & \multicolumn{2}{c}{$1{,}345{,}200$} \\
Pseudo $R^2$ & \multicolumn{2}{c}{$0.1267$} \\
\bottomrule
\end{tabular}
\vspace{2mm}
\parbox{0.745\textwidth}{\scriptsize\textit{Notes:} $^{***}p<0.01$, $^{**}p<0.05$, $^{*}p<0.10$. Logit on pairwise candidate differences; occupation-clustered standard errors in parentheses (177 clusters). All variables are differences between candidates $i$ and $j$ (e.g.\ Female $= \text{female}_i - \text{female}_j$). Each model evaluates 70{,}800 pairwise hiring decisions (177 occupations $\times$ 400 pairs). Ten matched instruct--base pairs (19 models, $19 \times 70{,}800 = 1{,}345{,}200$). Column~(1): coefficient in pre-trained models. Column~(2) ($\times$\,Instruct): additional effect in instruction-tuned models, estimated by interacting each regressor with an instruction-tuning indicator.}
\end{table}

%% file: table_intersectional_aggregate.tex
\begin{table}[h!]
\centering
\parbox{0.95\textwidth}{\caption{\textbf{\textbar{}\ Intersectional Effects: Aggregate Specification}}\label{tab:intersect_agg}}
\scriptsize
\begin{tabular}{lcc}
\toprule
 & (1) Base & (2) Instruct \\
\midrule
\textit{Panel A: Demographics} & & \\
Female & $0.0261^{***}$ $(0.0079)$ & $0.1305^{***}$ $(0.0121)$ \\
Black & $0.0240^{***}$ $(0.0073)$ & $0.0927^{***}$ $(0.0108)$ \\
Disabled & $-0.0293^{***}$ $(0.0073)$ & $-0.0572^{***}$ $(0.0111)$ \\[2pt]
\textit{Panel B: Intersectional demographics} & & \\
Female $\times$ Black & $-0.0006$ $(0.0091)$ & $-0.0168$ $(0.0120)$ \\
Female $\times$ Disabled & $-0.0007$ $(0.0097)$ & $-0.0072$ $(0.0132)$ \\
Black $\times$ Disabled & $-0.0099$ $(0.0081)$ & $-0.0110$ $(0.0127)$ \\[2pt]
\textit{Panel C: Controls} & & \\
Education & $0.0214^{***}$ $(0.0028)$ & $0.1220^{***}$ $(0.0116)$ \\
Work experience & $0.0938^{***}$ $(0.0028)$ & $0.5223^{***}$ $(0.0162)$ \\
GPA & $0.1422^{***}$ $(0.0030)$ & $0.4665^{***}$ $(0.0108)$ \\
Technology skills & $0.0881^{***}$ $(0.0071)$ & $0.4794^{***}$ $(0.0245)$ \\
Skill avg. & $0.0742^{***}$ $(0.0062)$ & $0.3386^{***}$ $(0.0182)$ \\
Position & $-0.0023$ $(0.0031)$ & $-0.0068$ $(0.0044)$ \\
\midrule
Observations & $637{,}200$ & $1{,}416{,}000$ \\
Pseudo $R^2$ & $-0.0030$ & $0.2886$ \\
\bottomrule
\end{tabular}
\vspace{2mm}
\parbox{0.879\textwidth}{\scriptsize\textit{Notes:} $^{***}p<0.01$, $^{**}p<0.05$, $^{*}p<0.10$. Logit on pairwise candidate differences; occupation-clustered standard errors in parentheses (177 clusters). All variables are differences between candidates $i$ and $j$ (e.g.\ Female $= \text{female}_i - \text{female}_j$). Each model evaluates 70{,}800 pairwise hiring decisions (177 occupations $\times$ 400 pairs). Column~(1): 9 pre-trained models ($9 \times 70{,}800 = 637{,}200$). Column~(2): 20 instruction-tuned models ($20 \times 70{,}800 = 1{,}416{,}000$). Intersectional terms (e.g.\ Female $\times$ Black) are differences in the product of two demographic indicators: $\text{female}_i \times \text{black}_i - \text{female}_j \times \text{black}_j$.}
\end{table}

%% file: table_cross_occupation.tex
\begin{table}[h!]
\centering
\parbox{0.95\textwidth}{\caption{\textbf{\textbar{}\ Cross-Occupation Variation}}\label{tab:cross_occ}}
\scriptsize
\begin{tabular}{lcc}
\toprule
 & (1) Base & (2) Instruct \\
\midrule
\textit{Panel A: Demographics} & & \\
Female & $-0.0013$ $(0.0103)$ & $0.1004^{***}$ $(0.0149)$ \\
Black & $0.0100$ $(0.0084)$ & $0.0629^{***}$ $(0.0158)$ \\
Disabled & $-0.0338^{***}$ $(0.0110)$ & $-0.0374^{**}$ $(0.0144)$ \\[2pt]
\textit{Panel B: Controls} & & \\
Education & $0.0214^{***}$ $(0.0027)$ & $0.1262^{***}$ $(0.0102)$ \\
Work experience & $0.0992^{***}$ $(0.0032)$ & $0.5708^{***}$ $(0.0151)$ \\
GPA & $0.1424^{***}$ $(0.0030)$ & $0.4749^{***}$ $(0.0103)$ \\
Technology skills & $0.0884^{***}$ $(0.0076)$ & $0.4760^{***}$ $(0.0263)$ \\
Skill avg. & $0.0735^{***}$ $(0.0061)$ & $0.3418^{***}$ $(0.0178)$ \\
Position & $-0.0024$ $(0.0031)$ & $-0.0079^{*}$ $(0.0044)$ \\[2pt]
\textit{Panel C: $\times$ Log hourly wage} & & \\
Female $\times$ Wage & $0.0045$ $(0.0043)$ & $0.0230^{***}$ $(0.0063)$ \\
Black $\times$ Wage & $0.0076$ $(0.0051)$ & $0.0144^{*}$ $(0.0073)$ \\
Disabled $\times$ Wage & $-0.0020$ $(0.0053)$ & $-0.0106$ $(0.0077)$ \\
Education $\times$ Wage & $0.0034$ $(0.0024)$ & $0.0007$ $(0.0096)$ \\
Work exp.\ $\times$ Wage & $-0.0152^{***}$ $(0.0029)$ & $-0.1110^{***}$ $(0.0106)$ \\
GPA $\times$ Wage & $0.0001$ $(0.0030)$ & $0.0255^{**}$ $(0.0126)$ \\
Tech skills $\times$ Wage & $-0.0059$ $(0.0073)$ & $-0.0028$ $(0.0248)$ \\
Skill avg.\ $\times$ Wage & $0.0096$ $(0.0060)$ & $0.0318^{*}$ $(0.0166)$ \\[2pt]
\textit{Panel D: $\times$ Employment} & & \\
Female $\times$ Employment & $0.0029$ $(0.0042)$ & $0.0106^{*}$ $(0.0060)$ \\
Black $\times$ Employment & $-0.0058$ $(0.0038)$ & $0.0022$ $(0.0061)$ \\
Disabled $\times$ Employment & $-0.0039$ $(0.0043)$ & $-0.0056$ $(0.0058)$ \\
Education $\times$ Employment & $-0.0045^{*}$ $(0.0026)$ & $-0.0244^{*}$ $(0.0125)$ \\
Work exp.\ $\times$ Employment & $0.0002$ $(0.0029)$ & $-0.0178$ $(0.0122)$ \\
GPA $\times$ Employment & $-0.0045$ $(0.0032)$ & $-0.0298$ $(0.0196)$ \\
Tech skills $\times$ Employment & $0.0097$ $(0.0073)$ & $-0.0018$ $(0.0267)$ \\
Skill avg.\ $\times$ Employment & $-0.0097^{*}$ $(0.0050)$ & $0.0106$ $(0.0158)$ \\[2pt]
\textit{Panel E: Demographic $\times$ Representation} & & \\
Female $\times$ Female representation & $0.0528^{***}$ $(0.0172)$ & $0.0381$ $(0.0240)$ \\
Black $\times$ Black representation & $0.0160$ $(0.0167)$ & $0.0333$ $(0.0271)$ \\
Disabled $\times$ Disability representation & $-0.0018$ $(0.0190)$ & $-0.0580^{**}$ $(0.0246)$ \\
\midrule
Observations & $637{,}200$ & $1{,}416{,}000$ \\
Pseudo $R^2$ & $-0.0024$ & $0.2961$ \\
\bottomrule
\end{tabular}
\vspace{2mm}
\parbox{0.915\textwidth}{\scriptsize\textit{Notes:} $^{***}p<0.01$, $^{**}p<0.05$, $^{*}p<0.10$. Logit on pairwise candidate differences; occupation-clustered standard errors in parentheses (177 clusters). All variables are differences between candidates $i$ and $j$ (e.g.\ Female $= \text{female}_i - \text{female}_j$). Each model evaluates 70{,}800 pairwise hiring decisions (177 occupations $\times$ 400 pairs). Column~(1): 9 pre-trained models ($9 \times 70{,}800 = 637{,}200$). Column~(2): 20 instruction-tuned models ($20 \times 70{,}800 = 1{,}416{,}000$). Demographic$\times$occupation interactions (e.g.\ Female $\times$ Wage) are the product of the candidate-level demographic difference and the occupation-level characteristic ($z$-scored log hourly wage, $z$-scored log employment, or within-sample percentile rank of group-specific representation rate). Control$\times$occupation interactions are defined analogously.}
\end{table}

%% file: table_cross_occupation_alignment.tex
\begin{table}[h!]
\centering
\parbox{0.95\textwidth}{\caption{\textbf{\textbar{}\ Alignment Effects: Cross-Occupation Variation}}\label{tab:cross_occ_align}}
\scriptsize
\begin{tabular}{lcc}
\toprule
 & (1) Base & (2) $\times$ Instruct \\
\midrule
\textit{Panel A: Demographics} & & \\
Female & $-0.0022$ $(0.0103)$ & $0.1112^{***}$ $(0.0149)$ \\
Black & $0.0101$ $(0.0084)$ & $0.0666^{***}$ $(0.0179)$ \\
Disabled & $-0.0314^{***}$ $(0.0110)$ & $-0.0183$ $(0.0164)$ \\[2pt]
\textit{Panel B: Controls} & & \\
Education & $0.0214^{***}$ $(0.0027)$ & $0.0783^{***}$ $(0.0075)$ \\
Work experience & $0.0992^{***}$ $(0.0032)$ & $0.3377^{***}$ $(0.0099)$ \\
GPA & $0.1424^{***}$ $(0.0030)$ & $0.3610^{***}$ $(0.0081)$ \\
Technology skills & $0.0883^{***}$ $(0.0076)$ & $0.2019^{***}$ $(0.0179)$ \\
Skill avg. & $0.0736^{***}$ $(0.0061)$ & $0.2331^{***}$ $(0.0138)$ \\
Position & $-0.0025$ $(0.0031)$ & $-0.0067$ $(0.0047)$ \\[2pt]
\textit{Panel C: $\times$ Log hourly wage} & & \\
Female $\times$ Wage & $0.0046$ $(0.0043)$ & $0.0323^{***}$ $(0.0060)$ \\
Black $\times$ Wage & $0.0075$ $(0.0051)$ & $0.0073$ $(0.0082)$ \\
Disabled $\times$ Wage & $-0.0028$ $(0.0053)$ & $-0.0062$ $(0.0095)$ \\
Education $\times$ Wage & $0.0034$ $(0.0024)$ & $-0.0009$ $(0.0071)$ \\
Work exp.\ $\times$ Wage & $-0.0152^{***}$ $(0.0029)$ & $-0.0693^{***}$ $(0.0069)$ \\
GPA $\times$ Wage & $0.0000$ $(0.0030)$ & $0.0151$ $(0.0092)$ \\
Tech skills $\times$ Wage & $-0.0060$ $(0.0073)$ & $0.0126$ $(0.0177)$ \\
Skill avg.\ $\times$ Wage & $0.0095$ $(0.0060)$ & $0.0211$ $(0.0129)$ \\[2pt]
\textit{Panel D: $\times$ Employment} & & \\
Female $\times$ Employment & $0.0029$ $(0.0041)$ & $0.0141^{**}$ $(0.0068)$ \\
Black $\times$ Employment & $-0.0058$ $(0.0038)$ & $0.0098$ $(0.0069)$ \\
Disabled $\times$ Employment & $-0.0039$ $(0.0043)$ & $0.0026$ $(0.0083)$ \\
Education $\times$ Employment & $-0.0045^{*}$ $(0.0026)$ & $-0.0170^{*}$ $(0.0098)$ \\
Work exp.\ $\times$ Employment & $0.0002$ $(0.0029)$ & $-0.0143^{*}$ $(0.0080)$ \\
GPA $\times$ Employment & $-0.0045$ $(0.0032)$ & $-0.0290^{**}$ $(0.0143)$ \\
Tech skills $\times$ Employment & $0.0093$ $(0.0073)$ & $-0.0056$ $(0.0171)$ \\
Skill avg.\ $\times$ Employment & $-0.0098^{*}$ $(0.0050)$ & $0.0195$ $(0.0137)$ \\[2pt]
\textit{Panel E: Demographic $\times$ Representation} & & \\
Female $\times$ Female representation & $0.0543^{***}$ $(0.0172)$ & $-0.0171$ $(0.0248)$ \\
Black $\times$ Black representation & $0.0157$ $(0.0167)$ & $0.0223$ $(0.0308)$ \\
Disabled $\times$ Disability representation & $-0.0063$ $(0.0190)$ & $-0.0688^{**}$ $(0.0283)$ \\
\midrule
Observations & \multicolumn{2}{c}{$1{,}345{,}200$} \\
Pseudo $R^2$ & \multicolumn{2}{c}{$0.1303$} \\
\bottomrule
\end{tabular}
\vspace{2mm}
\parbox{0.915\textwidth}{\scriptsize\textit{Notes:} $^{***}p<0.01$, $^{**}p<0.05$, $^{*}p<0.10$. Logit on pairwise candidate differences; occupation-clustered standard errors in parentheses (177 clusters). All variables are differences between candidates $i$ and $j$ (e.g.\ Female $= \text{female}_i - \text{female}_j$). Each model evaluates 70{,}800 pairwise hiring decisions (177 occupations $\times$ 400 pairs). Ten matched instruct--base pairs (19 models, $19 \times 70{,}800 = 1{,}345{,}200$). Column~(1): coefficient in pre-trained models. Column~(2) ($\times$\,Instruct): additional effect in instruction-tuned models, estimated by interacting each regressor with an instruction-tuning indicator. Demographic$\times$occupation terms are defined as in Extended Data Table~4.}
\end{table}

%% file: table_differential_returns.tex
\begin{table}[h!]
\centering
\parbox{0.95\textwidth}{\caption{\textbf{\textbar{}\ Differential Returns Specification}}\label{tab:diff_returns}}
\scriptsize
\begin{tabular}{lcc}
\toprule
 & (1) Base & (2) Instruct \\
\midrule
\textit{Panel A: Demographics} & & \\
Female & $-0.0184$ $(0.0244)$ & $-0.0078$ $(0.0375)$ \\
Black & $-0.0047$ $(0.0257)$ & $0.0262$ $(0.0450)$ \\
Disabled & $-0.0288$ $(0.0280)$ & $-0.1276^{***}$ $(0.0399)$ \\[2pt]
\textit{Panel B: Controls} & & \\
Education & $0.0195^{***}$ $(0.0030)$ & $0.1153^{***}$ $(0.0120)$ \\
Work experience & $0.0940^{***}$ $(0.0031)$ & $0.5131^{***}$ $(0.0164)$ \\
GPA & $0.1423^{***}$ $(0.0052)$ & $0.4691^{***}$ $(0.0119)$ \\
Technology skills & $0.0779^{***}$ $(0.0102)$ & $0.4669^{***}$ $(0.0278)$ \\
Skill avg. & $0.0360^{***}$ $(0.0091)$ & $0.1918^{***}$ $(0.0229)$ \\
Position & $-0.0045$ $(0.0087)$ & $0.0030$ $(0.0121)$ \\[2pt]
\textit{Panel C: Differential returns (female)} & & \\
Education $\times$ Female & $0.0025$ $(0.0016)$ & $0.0056^{**}$ $(0.0027)$ \\
Work exp.\ $\times$ Female & $0.0016$ $(0.0015)$ & $0.0209^{***}$ $(0.0029)$ \\
GPA $\times$ Female & $0.0029$ $(0.0047)$ & $0.0086$ $(0.0065)$ \\
Tech skills $\times$ Female & $0.0179^{*}$ $(0.0097)$ & $-0.0006$ $(0.0162)$ \\
Skill avg.\ $\times$ Female & $0.0309^{***}$ $(0.0107)$ & $0.1289^{***}$ $(0.0152)$ \\
Position $\times$ Female & $-0.0043$ $(0.0097)$ & $-0.0044$ $(0.0126)$ \\[2pt]
\textit{Panel D: Differential returns (Black)} & & \\
Education $\times$ Black & $0.0015$ $(0.0017)$ & $0.0024$ $(0.0031)$ \\
Work exp.\ $\times$ Black & $0.0004$ $(0.0017)$ & $0.0090^{***}$ $(0.0029)$ \\
GPA $\times$ Black & $-0.0023$ $(0.0046)$ & $-0.0032$ $(0.0055)$ \\
Tech skills $\times$ Black & $-0.0052$ $(0.0091)$ & $0.0119$ $(0.0144)$ \\
Skill avg.\ $\times$ Black & $0.0207^{**}$ $(0.0101)$ & $0.1179^{***}$ $(0.0167)$ \\
Position $\times$ Black & $0.0044$ $(0.0090)$ & $-0.0109$ $(0.0147)$ \\[2pt]
\textit{Panel E: Differential returns (disabled)} & & \\
Education $\times$ Disabled & $-0.0003$ $(0.0019)$ & $0.0056^{*}$ $(0.0028)$ \\
Work exp.\ $\times$ Disabled & $-0.0023$ $(0.0014)$ & $-0.0081^{***}$ $(0.0023)$ \\
GPA $\times$ Disabled & $-0.0010$ $(0.0048)$ & $-0.0091$ $(0.0062)$ \\
Tech skills $\times$ Disabled & $0.0079$ $(0.0091)$ & $0.0156$ $(0.0121)$ \\
Skill avg.\ $\times$ Disabled & $0.0447^{***}$ $(0.0113)$ & $0.1376^{***}$ $(0.0184)$ \\
Position $\times$ Disabled & $0.0044$ $(0.0092)$ & $-0.0054$ $(0.0136)$ \\
\midrule
Observations & $637{,}200$ & $1{,}416{,}000$ \\
Pseudo $R^2$ & $-0.0029$ & $0.2896$ \\
\bottomrule
\end{tabular}
\vspace{2mm}
\parbox{0.875\textwidth}{\scriptsize\textit{Notes:} $^{***}p<0.01$, $^{**}p<0.05$, $^{*}p<0.10$. Logit on pairwise candidate differences; occupation-clustered standard errors in parentheses (177 clusters). All variables are differences between candidates $i$ and $j$ (e.g.\ Female $= \text{female}_i - \text{female}_j$). Each model evaluates 70{,}800 pairwise hiring decisions (177 occupations $\times$ 400 pairs). Column~(1): 9 pre-trained models ($9 \times 70{,}800 = 637{,}200$). Column~(2): 20 instruction-tuned models ($20 \times 70{,}800 = 1{,}416{,}000$). Qualification$\times$demographic interactions (e.g.\ Work exp.\ $\times$ Female) capture differential returns: $\text{work\_exp}_i \times \text{female}_i - \text{work\_exp}_j \times \text{female}_j$.}
\end{table}

%% file: table_differential_returns_alignment.tex
\begin{table}[h!]
\centering
\parbox{0.95\textwidth}{\caption{\textbf{\textbar{}\ Alignment Effects: Differential Returns Specification}}\label{tab:diff_returns_align}}
\scriptsize
\begin{tabular}{lcc}
\toprule
 & (1) Base & (2) $\times$ instruct \\
\midrule
\textit{Panel A: Demographics} & & \\
Female & $-0.0182$ $(0.0244)$ & $-0.0292$ $(0.0438)$ \\
Black & $-0.0036$ $(0.0256)$ & $0.0222$ $(0.0518)$ \\
Disabled & $-0.0298$ $(0.0280)$ & $-0.0912^{*}$ $(0.0501)$ \\[2pt]
\textit{Panel B: Controls} & & \\
Education & $0.0196^{***}$ $(0.0030)$ & $0.0695^{***}$ $(0.0092)$ \\
Work experience & $0.0941^{***}$ $(0.0031)$ & $0.2935^{***}$ $(0.0103)$ \\
GPA & $0.1424^{***}$ $(0.0052)$ & $0.3524^{***}$ $(0.0098)$ \\
Technology skills & $0.0767^{***}$ $(0.0102)$ & $0.1991^{***}$ $(0.0200)$ \\
Skill avg. & $0.0360^{***}$ $(0.0091)$ & $0.1354^{***}$ $(0.0187)$ \\
Position & $-0.0043$ $(0.0087)$ & $0.0085$ $(0.0128)$ \\[2pt]
\textit{Panel C: Differential returns (Female)} & & \\
Education $\times$ Female & $0.0025$ $(0.0016)$ & $0.0058^{*}$ $(0.0031)$ \\
Work exp.\ $\times$ Female & $0.0016$ $(0.0015)$ & $0.0210^{***}$ $(0.0027)$ \\
GPA $\times$ Female & $0.0029$ $(0.0047)$ & $0.0087$ $(0.0072)$ \\
Tech skills $\times$ Female & $0.0187^{*}$ $(0.0097)$ & $-0.0171$ $(0.0163)$ \\
Skill avg.\ $\times$ Female & $0.0310^{***}$ $(0.0107)$ & $0.0813^{***}$ $(0.0168)$ \\
Position $\times$ Female & $-0.0057$ $(0.0097)$ & $0.0037$ $(0.0148)$ \\[2pt]
\textit{Panel D: Differential returns (Black)} & & \\
Education $\times$ Black & $0.0014$ $(0.0017)$ & $0.0025$ $(0.0037)$ \\
Work exp.\ $\times$ Black & $0.0004$ $(0.0017)$ & $0.0098^{***}$ $(0.0030)$ \\
GPA $\times$ Black & $-0.0023$ $(0.0046)$ & $-0.0023$ $(0.0066)$ \\
Tech skills $\times$ Black & $-0.0046$ $(0.0091)$ & $0.0200$ $(0.0167)$ \\
Skill avg.\ $\times$ Black & $0.0208^{**}$ $(0.0101)$ & $0.0869^{***}$ $(0.0175)$ \\
Position $\times$ Black & $0.0049$ $(0.0090)$ & $-0.0199$ $(0.0158)$ \\[2pt]
\textit{Panel E: Differential returns (Disabled)} & & \\
Education $\times$ Disabled & $-0.0002$ $(0.0019)$ & $0.0035$ $(0.0035)$ \\
Work exp.\ $\times$ Disabled & $-0.0023$ $(0.0014)$ & $-0.0036$ $(0.0028)$ \\
GPA $\times$ Disabled & $-0.0011$ $(0.0048)$ & $-0.0012$ $(0.0071)$ \\
Tech skills $\times$ Disabled & $0.0084$ $(0.0091)$ & $0.0151$ $(0.0134)$ \\
Skill avg.\ $\times$ Disabled & $0.0447^{***}$ $(0.0113)$ & $0.0799^{***}$ $(0.0177)$ \\
Position $\times$ Disabled & $0.0049$ $(0.0092)$ & $-0.0135$ $(0.0151)$ \\
\midrule
Observations & \multicolumn{2}{c}{$1{,}345{,}200$} \\
Pseudo $R^2$ & \multicolumn{2}{c}{$0.1273$} \\
\bottomrule
\end{tabular}
\vspace{2mm}
\parbox{0.865\textwidth}{\scriptsize\textit{Notes:} $^{***}p<0.01$, $^{**}p<0.05$, $^{*}p<0.10$. Logit on pairwise candidate differences; occupation-clustered standard errors in parentheses (177 clusters). All variables are differences between candidates $i$ and $j$ (e.g.\ Female $= \text{female}_i - \text{female}_j$). Each model evaluates 70{,}800 pairwise hiring decisions (177 occupations $\times$ 400 pairs). Ten matched instruct--base pairs (19 models, $19 \times 70{,}800 = 1{,}345{,}200$). Column~(1): coefficient in pre-trained models. Column~(2) ($\times$\,Instruct): additional effect in instruction-tuned models, estimated by interacting each regressor with an instruction-tuning indicator. Qualification$\times$demographic interactions are defined as in Extended Data Table~6.}
\end{table}

%% file: table_information_asymmetry.tex
\begin{table}[h!]
\centering
\parbox{0.95\textwidth}{\caption{\textbf{\textbar{}\ Information Asymmetry Specification}}\label{tab:info_asymmetry}}
\scriptsize
\begin{tabular}{lcc}
\toprule
 & (1) Base & (2) Instruct \\
\midrule
\textit{Panel A: Demographics} & & \\
Female & $0.0246^{***}$ $(0.0047)$ & $0.1364^{***}$ $(0.0068)$ \\
Black & $0.0191^{***}$ $(0.0051)$ & $0.0944^{***}$ $(0.0068)$ \\
Disabled & $-0.0345^{***}$ $(0.0045)$ & $-0.0715^{***}$ $(0.0064)$ \\[2pt]
\textit{Panel B: Controls} & & \\
Education & $0.0207^{***}$ $(0.0023)$ & $0.1332^{***}$ $(0.0084)$ \\
Work experience & $0.0640^{***}$ $(0.0019)$ & $0.4030^{***}$ $(0.0105)$ \\
GPA & $0.1337^{***}$ $(0.0028)$ & $0.4771^{***}$ $(0.0089)$ \\
Technology skills & $0.0938^{***}$ $(0.0070)$ & $0.4939^{***}$ $(0.0230)$ \\
Skill avg.\ (cond.) & $0.0969^{***}$ $(0.0066)$ & $0.4610^{***}$ $(0.0167)$ \\
Position & $-0.0003$ $(0.0038)$ & $-0.0008$ $(0.0043)$ \\[2pt]
\textit{Panel C: Information asymmetry ($v$)} & & \\
$v$ (work exp.\ absent) & $-0.7037^{***}$ $(0.0473)$ & $-3.6396^{***}$ $(0.3285)$ \\
$v \times$ Female & $-0.0022$ $(0.0091)$ & $-0.0400^{***}$ $(0.0151)$ \\
$v \times$ Black & $0.0008$ $(0.0103)$ & $-0.0191$ $(0.0184)$ \\
$v \times$ Disabled & $-0.0128$ $(0.0101)$ & $-0.0496^{***}$ $(0.0159)$ \\
$v \times$ Education & $0.0141^{***}$ $(0.0032)$ & $0.0946^{***}$ $(0.0237)$ \\
$v \times$ GPA & $0.0541^{***}$ $(0.0064)$ & $0.4398^{***}$ $(0.0443)$ \\
$v \times$ Tech skills & $-0.0438^{***}$ $(0.0147)$ & $0.0377$ $(0.0505)$ \\
$v \times$ Position & $-0.0011$ $(0.0099)$ & $-0.0093$ $(0.0132)$ \\
\midrule
Observations & $637{,}200$ & $1{,}416{,}000$ \\
Pseudo $R^2$ & $0.0076$ & $0.3679$ \\
\bottomrule
\end{tabular}
\vspace{2mm}
\parbox{0.865\textwidth}{\scriptsize\textit{Notes:} $^{***}p<0.01$, $^{**}p<0.05$, $^{*}p<0.10$. Logit on pairwise candidate differences; occupation-clustered standard errors in parentheses (177 clusters). All variables are differences between candidates $i$ and $j$ (e.g.\ Female $= \text{female}_i - \text{female}_j$). Each model evaluates 70{,}800 pairwise hiring decisions (177 occupations $\times$ 400 pairs). Column~(1): 9 pre-trained models ($9 \times 70{,}800 = 637{,}200$). Column~(2): 20 instruction-tuned models ($20 \times 70{,}800 = 1{,}416{,}000$). $v = \mathbf{1}[\text{work exp.} = 0]$ is an information-asymmetry indicator. $v \times$ demographic interactions (e.g.\ $v \times$ Female $= v_i \times \text{female}_i - v_j \times \text{female}_j$) test whether models penalise candidates differently by group when qualification signals are absent. Skill avg.\ (cond.) $= (1 - v_i) \times \overline{\text{skill}}_i - (1 - v_j) \times \overline{\text{skill}}_j$.}
\end{table}

%% file: table_information_asymmetry_alignment.tex
\begin{table}[h!]
\centering
\parbox{0.95\textwidth}{\caption{\textbf{\textbar{}\ Alignment Effects: Information Asymmetry Specification}}\label{tab:info_asymmetry_align}}
\scriptsize
\begin{tabular}{lcc}
\toprule
 & (1) Base & (2) $\times$ instruct \\
\midrule
\textit{Panel A: Demographics} & & \\
Female & $0.0247^{***}$ $(0.0047)$ & $0.1242^{***}$ $(0.0077)$ \\
Black & $0.0190^{***}$ $(0.0051)$ & $0.0971^{***}$ $(0.0083)$ \\
Disabled & $-0.0344^{***}$ $(0.0045)$ & $-0.0548^{***}$ $(0.0072)$ \\[2pt]
\textit{Panel B: Controls} & & \\
Education & $0.0207^{***}$ $(0.0023)$ & $0.0856^{***}$ $(0.0063)$ \\
Work experience & $0.0640^{***}$ $(0.0019)$ & $0.2370^{***}$ $(0.0067)$ \\
GPA & $0.1337^{***}$ $(0.0028)$ & $0.3724^{***}$ $(0.0079)$ \\
Technology skills & $0.0938^{***}$ $(0.0070)$ & $0.2147^{***}$ $(0.0157)$ \\
Skill avg.\ (cond.) & $0.0969^{***}$ $(0.0066)$ & $0.3139^{***}$ $(0.0129)$ \\
Position & $-0.0004$ $(0.0038)$ & $0.0014$ $(0.0052)$ \\[2pt]
\textit{Panel C: Information asymmetry ($v$)} & & \\
$v$ (work exp.\ absent) & $-0.7036^{***}$ $(0.0473)$ & $-1.8800^{***}$ $(0.2309)$ \\
$v \times$ Female & $-0.0024$ $(0.0091)$ & $-0.0666^{***}$ $(0.0173)$ \\
$v \times$ Black & $0.0008$ $(0.0103)$ & $-0.0541^{**}$ $(0.0211)$ \\
$v \times$ Disabled & $-0.0125$ $(0.0101)$ & $-0.0716^{***}$ $(0.0199)$ \\
$v \times$ Education & $0.0141^{***}$ $(0.0032)$ & $0.0383^{**}$ $(0.0168)$ \\
$v \times$ GPA & $0.0541^{***}$ $(0.0064)$ & $0.2980^{***}$ $(0.0324)$ \\
$v \times$ Tech skills & $-0.0451^{***}$ $(0.0147)$ & $0.0152$ $(0.0415)$ \\
$v \times$ Position & $-0.0009$ $(0.0099)$ & $-0.0126$ $(0.0161)$ \\
\midrule
Observations & \multicolumn{2}{c}{$1{,}345{,}200$} \\
Pseudo $R^2$ & \multicolumn{2}{c}{$0.1688$} \\
\bottomrule
\end{tabular}
\vspace{2mm}
\parbox{0.865\textwidth}{\scriptsize\textit{Notes:} $^{***}p<0.01$, $^{**}p<0.05$, $^{*}p<0.10$. Logit on pairwise candidate differences; occupation-clustered standard errors in parentheses (177 clusters). All variables are differences between candidates $i$ and $j$ (e.g.\ Female $= \text{female}_i - \text{female}_j$). Each model evaluates 70{,}800 pairwise hiring decisions (177 occupations $\times$ 400 pairs). Ten matched instruct--base pairs (19 models, $19 \times 70{,}800 = 1{,}345{,}200$). Column~(1): coefficient in pre-trained models. Column~(2) ($\times$\,Instruct): additional effect in instruction-tuned models, estimated by interacting each regressor with an instruction-tuning indicator. $v$, Skill avg.\ (cond.), and $v \times$ demographic interactions are defined as in Extended Data Table~8.}
\end{table}

%% file: table_nonparametric_base.tex
\begin{table}[h!]
\centering
\parbox{0.85\textwidth}{\caption{\textbf{\textbar{}\ Nonparametric Robustness: Pre-Trained Models}}\label{tab:nonparam_base}}
\scriptsize
\begin{tabular}{lccc}
\toprule
 & (1) Logistic & (2) XGB & (3) RF \\
\midrule
\textit{Panel A: Demographics} & & & \\
Female & $0.0255^{***}$ $(0.0044)$ & $0.0316^{***}$ $(0.0053)$ & $0.0185^{***}$ $(0.0030)$ \\
Black & $0.0187^{***}$ $(0.0044)$ & $0.0100^{*}$ $(0.0050)$ & $0.0043^{*}$ $(0.0024)$ \\
Disabled & $-0.0345^{***}$ $(0.0042)$ & $-0.0319^{***}$ $(0.0056)$ & $-0.0173^{***}$ $(0.0030)$ \\[2pt]
\textit{Panel B: Controls} & & & \\
Education & $0.0214^{***}$ $(0.0028)$ & $0.0176^{***}$ $(0.0043)$ & $0.0101^{***}$ $(0.0016)$ \\
Work experience & $0.0939^{***}$ $(0.0028)$ & $0.1346^{***}$ $(0.0024)$ & $0.1301^{***}$ $(0.0019)$ \\
GPA & $0.1422^{***}$ $(0.0030)$ & $0.1170^{***}$ $(0.0143)$ & $0.0917^{***}$ $(0.0040)$ \\
Technology skills & $0.0881^{***}$ $(0.0071)$ & $0.0395^{***}$ $(0.0056)$ & $0.0150^{***}$ $(0.0032)$ \\
Skill avg. & $0.0741^{***}$ $(0.0062)$ & $0.0865^{***}$ $(0.0144)$ & $0.0468^{***}$ $(0.0069)$ \\
Position & $-0.0023$ $(0.0031)$ & $-0.0037$ $(0.0045)$ & $-0.0006$ $(0.0017)$ \\
\midrule
Observations & \multicolumn{3}{c}{$637{,}200$} \\
\bottomrule
\end{tabular}
\vspace{2mm}
\parbox{0.985\textwidth}{\scriptsize\textit{Notes:} $^{***}p<0.01$, $^{**}p<0.05$, $^{*}p<0.1$. Standard errors in parentheses. Column~(1): logistic regression coefficients from Extended Data Table~1 (pre-trained models); occupation-clustered standard errors. Column~(2): XGBoost (XGB) estimates; standard errors from 1{,}000 occupation-level cluster bootstrap iterations. Column~(3): random forest (RF) estimates; standard errors from 1{,}000 occupation-level cluster bootstrap iterations. Columns~(2)--(3) are computed via counterfactual prediction (Methods~III). $N = 637{,}200$ pairwise decisions from 9 pre-trained models across 177 occupations.}
\end{table}

%% file: table_nonparametric_instruct.tex
\begin{table}[h!]
\centering
\parbox{0.85\textwidth}{\caption{\textbf{\textbar{}\ Nonparametric Robustness: Instruction-Tuned Models}}\label{tab:nonparam_instruct}}
\scriptsize
\begin{tabular}{lccc}
\toprule
 & (1) Logistic & (2) XGB & (3) RF \\
\midrule
\textit{Panel A: Demographics} & & & \\
Female & $0.1186^{***}$ $(0.0065)$ & $0.1379^{***}$ $(0.0081)$ & $0.0764^{***}$ $(0.0045)$ \\
Black & $0.0788^{***}$ $(0.0071)$ & $0.0778^{***}$ $(0.0069)$ & $0.0326^{***}$ $(0.0032)$ \\
Disabled & $-0.0663^{***}$ $(0.0059)$ & $-0.0998^{***}$ $(0.0067)$ & $-0.0519^{***}$ $(0.0035)$ \\[2pt]
\textit{Panel B: Controls} & & & \\
Education & $0.1220^{***}$ $(0.0116)$ & $0.0822^{***}$ $(0.0167)$ & $0.0762^{***}$ $(0.0063)$ \\
Work experience & $0.5223^{***}$ $(0.0162)$ & $0.5435^{***}$ $(0.0052)$ & $0.4741^{***}$ $(0.0108)$ \\
GPA & $0.4665^{***}$ $(0.0108)$ & $0.4096^{***}$ $(0.0252)$ & $0.3055^{***}$ $(0.0086)$ \\
Technology skills & $0.4794^{***}$ $(0.0245)$ & $0.3205^{***}$ $(0.0175)$ & $0.2029^{***}$ $(0.0134)$ \\
Skill avg. & $0.3386^{***}$ $(0.0182)$ & $0.3266^{***}$ $(0.0255)$ & $0.1678^{***}$ $(0.0132)$ \\
Position & $-0.0068$ $(0.0044)$ & $-0.0001$ $(0.0058)$ & $-0.0004$ $(0.0022)$ \\
\midrule
Observations & \multicolumn{3}{c}{$1{,}416{,}000$} \\
\bottomrule
\end{tabular}
\vspace{2mm}
\parbox{0.985\textwidth}{\scriptsize\textit{Notes:} $^{***}p<0.01$, $^{**}p<0.05$, $^{*}p<0.1$. Standard errors in parentheses. Column~(1): logistic regression coefficients from Extended Data Table~1 (instruction-tuned models); occupation-clustered standard errors. Column~(2): XGBoost (XGB) estimates; standard errors from 1{,}000 occupation-level cluster bootstrap iterations. Column~(3): random forest (RF) estimates; standard errors from 1{,}000 occupation-level cluster bootstrap iterations. Columns~(2)--(3) are computed via counterfactual prediction (Methods~III). $N = 1{,}416{,}000$ pairwise decisions from 20 instruction-tuned models across 177 occupations.}
\end{table}